\documentclass[twocolumn,english,aps,superscriptaddress,amsmath,amssymb,floatfix]{revtex4-1}
\usepackage[colorlinks=true,citecolor=blue,linkcolor=magenta]{hyperref}

\usepackage[markup=blue, authormarkupposition=left]{changes} 

\usepackage{soul}
\usepackage[utf8]{inputenc}
\usepackage[english]{babel}
\usepackage{amsmath,amsfonts,amssymb}
\usepackage[T1]{fontenc}
\usepackage{url}

\usepackage{cancel}
\usepackage{amsmath}
\usepackage{amsfonts}
\usepackage{amssymb}

\usepackage{epstopdf}
\usepackage{graphicx}
\begin{document}

\title{Dual Laser Self-Injection Locking to a Micro Fabry-Perot for Low Phase Noise Millimeter-wave Generation}

%\title{Record Low Noise Millimeter-Wave Generation via \\Photonic SIL Lasers and Ultrastable Fabry-Perot Heterodyne}

\author{William Groman}
\affiliation{Electrical Computer \& Energy Engineering, University of Colorado, Boulder, CO, USA}
\affiliation{Department of Physics, University of Colorado, Boulder, Colorado, USA}

\author{Naijun Jin}
\affiliation{Department of Applied Physics, Yale University, New Haven, Connecticut, USA}

\author{Haotian Cheng}
\affiliation{Department of Applied Physics, Yale University, New Haven, Connecticut, USA}

\author{Dylan Meyer}
\affiliation{Electrical Computer \& Energy Engineering, University of Colorado, Boulder, CO, USA}
\affiliation{Department of Physics, University of Colorado, Boulder, Colorado, USA}

\author{Matthew Heyrich}
\affiliation{Electrical Computer \& Energy Engineering, University of Colorado, Boulder, CO, USA}
\affiliation{Department of Physics, University of Colorado, Boulder, Colorado, USA}

\author{Yifan Liu}
\affiliation{Department of Physics, University of Colorado, Boulder, Colorado, USA}
\affiliation{Time and Frequency Division, National Institute of Standards and Technology, Boulder, Colorado, USA}

\author{Alexander Lind}
\affiliation{Electrical Computer \& Energy Engineering, University of Colorado, Boulder, CO, USA}
\affiliation{Department of Physics, University of Colorado, Boulder, Colorado, USA}

\author{Charles A. McLemore}
\affiliation{Time and Frequency Division, National Institute of Standards and Technology, Boulder, Colorado, USA}

\author{Franklyn Quinlan}
\affiliation{Time and Frequency Division, National Institute of Standards and Technology, Boulder, Colorado, USA}
\affiliation{Electrical Computer \& Energy Engineering, University of Colorado, Boulder, CO, USA}

\author{Peter Rakich}
\affiliation{Department of Applied Physics, Yale University, New Haven, Connecticut, USA}

\author{Scott A. Diddams}
\affiliation{Electrical Computer \& Energy Engineering, University of Colorado, Boulder, CO, USA}
\affiliation{Department of Physics, University of Colorado, Boulder, Colorado, USA}

\begin{abstract}

Low-noise and accessible millimeter-wave sources are critical for emergent telecommunications, radar and sensing applications. Current limitations to realizing low-noise, deployable millimeter-wave systems include size, weight, and power (SWaP) requirements, along with complex operating principles. In this paper we provide a compact photonic implementation for generating low phase noise millimeter-waves, which significantly simplifies the architecture and reduces the volume compared to alternative approaches. Two commercial diode lasers are self-injection-locked to a micro-Fabry-Perot cavity, and their heterodyne provides low phase noise millimeter waves reaching -148 dBc/Hz at 1 MHz offset on a 111.45 GHz carrier. Phase noise characterization at such levels and frequencies poses unique challenges, and we further highlight the capabilities of optically-based measurement techniques. Our approach to millimter-wave generation can leverage advances in photonic integration for further miniaturization and packaging, thus providing a unique source of accessible, compact, and low-noise millimeter waves. 

\end{abstract}

\maketitle

\section*{Introduction}

Millimeter-wave and terahertz wave technologies have important applications in next generation telecommunications \cite{Nagatsuma2016,Lu2024,Lu2023,Dittmer2024}, radar and 3D~imaging \cite{Dong2022,Yi2023,Bai2023}, and bio-sensing \cite{Zhao2024,Chen2024,Zhang2023}. In communications systems, the higher frequency and associated modulation bandwidths, allow millimeter-waves to carry more data.
%For example, data rates  greater than 100 Gb/s have been projected (or achieved?, insert reference). 
Shorter wavelengths also allow for radar with sub-millimeter  spatial resolution \cite{Yi2023}. Important for both cases is decreased phase noise, which facilitates low bit-error rates, higher quadrature modulation capability, and higher radar and imaging fidelity \cite{Dittmer2024}. Addtionally, millimeter-waves provide unique opportunities in near-field imaging and human sensing (i.e. vital signs and human recognition/classification) \cite{Zhang2023, Chen2024}. Throughout this developing application landscape, cost effective and simple-to-implement millimeter-wave sources will enable impactful advances for a wider research community, and the general population. 

%Millimeter-wave carriers provide higher data transfer rate capabilities than the current standard of microwave communications. Additionally, their smaller wavelengths provide an improvement to sensing and radar resolution compared to microwave-based systems. 

%In this developing landscape, millimeter-wave sources that are both low-noise and compact are required to make these advancements widespread and available.  

Recent progress in photonically-generated millimeter-wave sources offer viable solutions to these requirements, with systems evolving from table-top volumes \cite{Fortier2016, Gan2015} to hybrid and integrated platforms \cite{Testumoto2021,Heffernan2024,Sun2024,Groman2024} %andhandheld photonic crystal oscillators \cite{Lia2023,Salek2023} 
in the past decade. Indeed, systems employing integrated lasers, microresonators, compact Fabry-Perot (FP) cavities, and microcombs, have achieved microwave and millimeter-wave signals with phase noise performance beyond commercial solutions, and comparable to bulk optic system performance \cite{Sun2024,Groman2024}. Still, many aspects of these systems require complex schemes based on high-bandwidth active servo control (e.g. Pound-Drever-Hall, PDH) and phase-lock loops. 

Alternatively, self-injection locking (SIL) \cite{Hadley1986} with chip-scale lasers and integrated optical cavities is a potential path around many of the complexities of active laser control. As advantages, SIL can employ inexpensive and noisy diode lasers and yields superior noise performance far from carrier, when compared to active servo control. 
% Optically derived mm-waves utilizing self-injection-locking schemes provide a way to forego these locking and power requirements.  
When implemented with recent advances in vacuum-sealed and integrable Fabry-Perot (FP) cavities \cite{Jin2022, Liu2024OPTICA}, spiral microresonators \cite{Li2021,LiuBlumenthal2022}, and fiber cavities \cite{Heffernan2024}, SIL lasers yield a compelling platform for integrated compact millimeter-wave generators. 

In this work, we demonstrate passive self-injection-locking of two diode lasers to a high finesse ($F~\approx$~400,000), wafer-fabricated micro-FP cavity for low-noise millimeter-wave generation. By self-injection-locking two lasers to the same micro-FP cavity, we leverage common mode rejection to suppress the noise of the resultant mm-wave beyond the noise of the individual lasers (i.e. the thermal noise limit of the cavity and fiber noise). The heterodyne of the two lasers that are self-injection-locked to two resonant modes of the same micro-FP cavity generates a 111.45 GHz millimeter-wave with phase noise as low as -148 dBc/Hz at 1 MHz offset frequency. This achievement represents a low-noise, compact and manufacturable, millimeter-wave reference system, which competes with state-of-the-art mm-wave generators and avoids the increase in phase noise inherent in active phase locking. 

\begin{figure}[]

  \centering
  \includegraphics[width= \columnwidth]{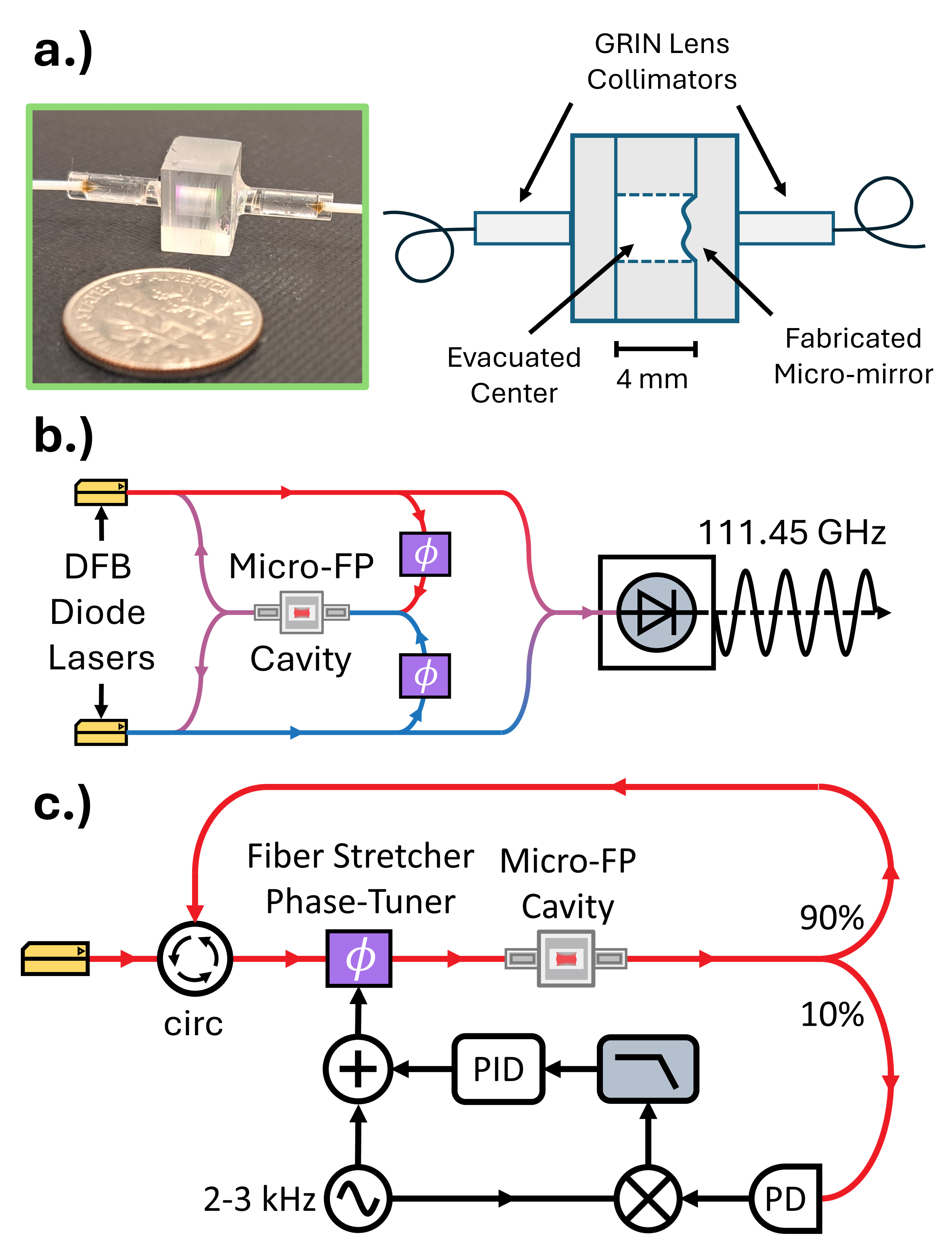}

\caption{a.) Photo of the micro-FP cavity in green box (left). Pictorial representation of the micro-FP cavity (right). b.) Simplified schematic of dual-SIL system. Phase controls are indicated with purple boxes. c.) Schematic of dither-locking system. %c.) Dual-tone delayed-self-heterodyne measurement system.
}
  \label{1}
 \vspace{-.32 cm}
\end{figure}

\section*{Concept \& Experiment}

%Inserted Frank's text
The micro-FP cavity used in this work is shown in Fig. 1a. Cavity fabrication starts with an array of mirrors lithographically patterned and etched on a 2 mm thick, 50.4 mm diameter ULE wafer, as described in \cite{Jin2022}. A 4 mm thick, 50.4 mm diameter ULE wafer with an array of bore holes matching the mirror array locations is optically contact bonded to the mirror array. On the opposite end of the 4 mm thick wafer, a 2 mm thick flat mirror with matching diameter is also optically contact bonded. Importantly, the wafers are bonded together in a vacuum environment \cite{Liu2024OPTICA}, creating an array of vacuum-sealed FP cavities. Individual cavities are then diced out of the array. For the cavity used here, two fiber collimators are aligned and glued to the ends of the cavity, allowing light coupling without the need for free-space alignment.
%A 4 mm long micro-FP cavity is fabricated on a wafer and diced \cite{Jin2022}, then two fiber collimators are aligned and glued to the ends of the cavity, shown in Fig.~\ref{1}a.). This allows light that is resonant with the cavity to be coupled in and out of the cavity without the need for free-space alignment. 
Two distributed feedback (DFB) diode lasers (linewidth < 500 kHz) are then frequency-tuned, via current, to two unique longitudinal modes of the micro-FP cavity that are separated by three free spectral ranges (FSR $\approx$ 37.15 GHz). The lasers are self-injection-locked (SIL) to the micro-FP cavity by feeding the resonant, cavity-transmitted light back to the individual lasers, as pictured in Fig.~\ref{1}b. To enable the self-injection locking of the lasers to the micro-FP, the phase of each laser light is tuned using a piezo-controlled fiber stretcher (purple boxes in Fig.~\ref{1}b.). Self-injection-locking allows the commercial diode lasers to inherit the stability of the micro-FP, applying a flat gain up to the cavity linewidth, by seeding the lasers with cavity-transmitted (i.e. filtered) light. The light from each laser is split before the individual phase controllers, combined onto one fiber and heterodyned on a millimeter-wave photodetector, producing  0.1 mW at the 111.45 GHz millimeter-wave carrier signal.

Long-term self-injection-locking is limited by fluctuations in fiber and cavity path lengths. To overcome this we implemented low-bandwidth  feedback to each laser, based on a cavity transmission dither-locking scheme, pictured in Fig.~\ref{1}c.). We dither the two phase controllers at unique frequencies (2 kHz and 3 kHz), photodetect the cavity-transmitted light and demodulate, providing an error signal that is filtered and conditioned before being applied to the voltage of each phase tuner. This allows us to keep the lasers self-injection-locked for up to an hour (5-10 times longer than without active feedback) while also reducing low offset frequency phase noise incurred by fiber/acoustic noise. Improved robustnes of SIL is expected with full integration of the optical circuit and cavity \cite{ChengYale2024}.

The phase noise of the resulting millimeter-wave was measured using cross-correlation techniques \cite{Rubiola2010}. %Insert Frank's text
Cross-correlation involves comparing the signal of interest against two phase-independent synthesizers; this method allows one to reject the noise of the two synthesizers, leaving only the noise of the signal under test, by averaging repeated measurements. The amount of noise rejection provided by cross-correlation scales with $1/\sqrt m$, where $m$ is the number of correlations, or measurements.%
%Cross-correlation involves comparing the signal of interest against two phase-independent synthesizers, which "averages out" the uncorrelated noise of the two synthesizers and thereby achieves sensitivity below the phase noise levels of the individual synthesizers. The cross-correlation measurement noise floor scales with $\sqrt{1/m}$, where m is the number of correlations, or measurements. 

\begin{figure*}[]
  \includegraphics[width=\linewidth]{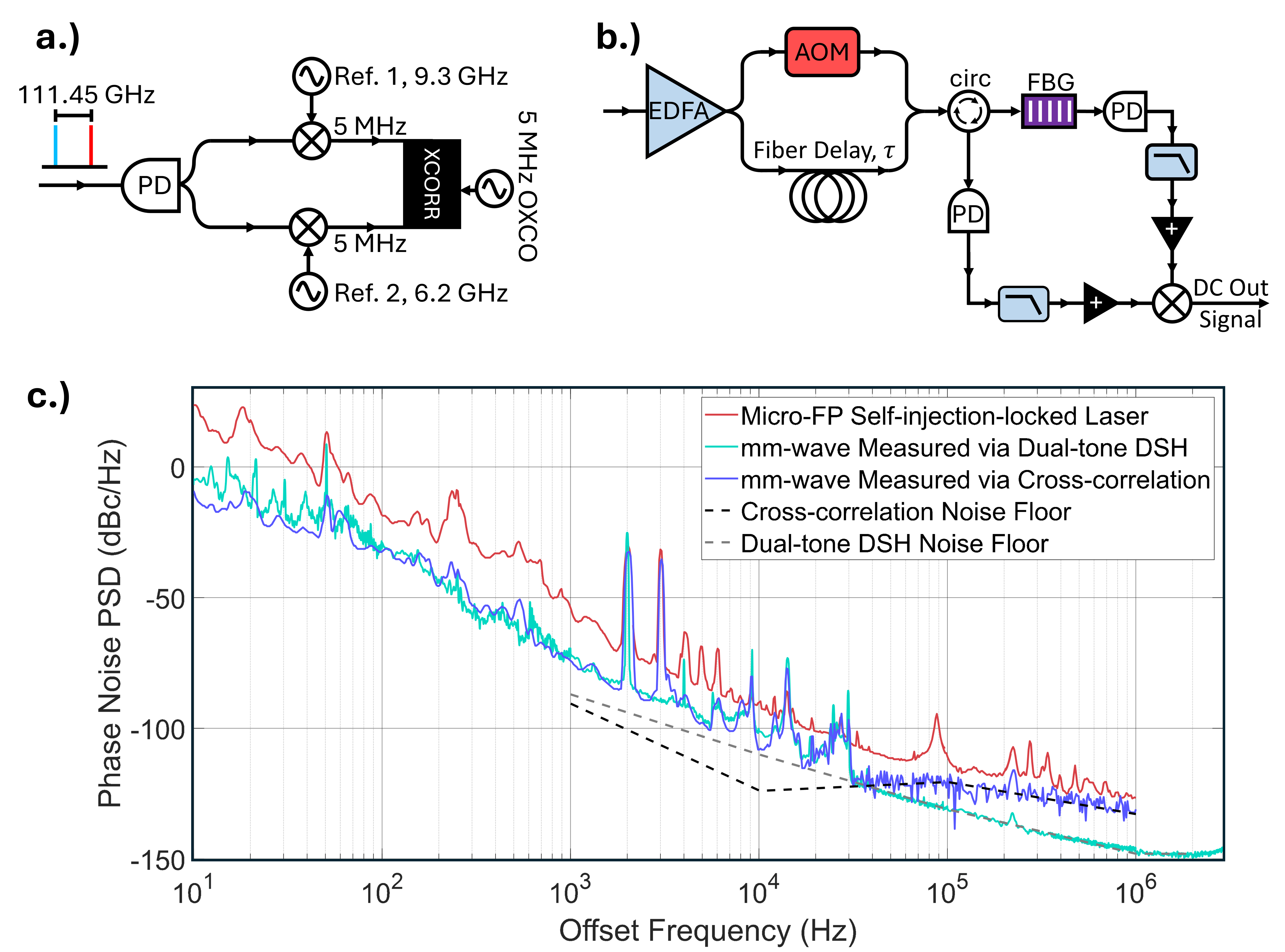}
\caption{a.) Schematic of cross-correlation system using two harmonic mixers (N = 12, N = 18) fed into a phase noise analyser (labelled XCORR) and referenced to an external crystal oscillator (OCXO). b.)  Schematic of dual-tone delayed self-heterodyne measurement system. c.) Plot of cross-correlation measurement of dual-SIL millimeter-wave heterodyne (blue) versus dual-tone delayed-self-heterodyne measurement of dual-SIL millimeter-wave heterodyne (teal). Single laser (self-injection-locked to micro-FP) phase noise (red) is measured using heterodyne with a stable laser and delayed-self-heterodyne with 55 m up to 6000 m fiber delays. In black dashed line, cross-correlation measurement noise floor is plotted; in grey the dual-tone DSH measurement noise floor is plotted.}
\label{2}
\end{figure*}

The cross-correlation measurement system is shown in Fig.~\ref{2}a. First, we split the photo-detected millimeter-wave signal and mix with the outputs of independent microwave references (9.3 GHz and 6.2 GHz) in two harmonic mixers. The mixers produce harmonic sidebands (N = 12, N = 18) that are approximately 5 MHz away from the optically derived millimeter-wave signal. The two 5 MHz down-converted signals are then cross-correlated by a phase noise analyzer that is referenced with a 5 MHz signal from a crystal oscillator. The resulting phase noise is shown in blue in Fig.~\ref{2}c. In the black dashed line, we plot the measurement limit of the cross-correlation measurement system after 40 minutes of averaging. We infer from this that the millimeter-wave phase noise is measurement-limited at offset frequencies above 30 kHz. 

To overcome this measurement limitation, we implemented a dual-tone delayed-self-heterodyne (DSH) measurement system based on \cite{Kwon2017} and shown in Fig.~\ref{2}b. First, both optical tones are combined in one fiber, amplified, and split along two paths: one path introduces a delay and the second path introduces a frequency shift of 100 MHz with an acousto-optic modulator. The now four tones are recombined in one fiber and split based on wavelength using a fiber Bragg grating (FBG). This allows us to split the light coming from each laser and photo-detect the 100 MHz heterodyned signals, similar to single-tone DSH. The electronic 100 MHz beats are low-pass filtered, amplified, then mixed together. The mixing stage allows noise that is correlated between the optical tones (i.e. from DSH and self-injection-locking to the same micro-FP) to cancel out. Lastly, the DC voltage noise power spectral density (PSD) from the output of the mixer is measured. 

A Hilbert transform is applied to the voltage noise PSD, based on the transfer function of DSH and the voltage-to-phase calibration of the measurement device. The transfer function of dual-tone DSH is similar to the transfer function of single-tone DSH, because both optical tones commonly sample the delay line, and acousto-optic modulator \cite{Kwon2017}:

\begin{equation}
\mathcal{L}_\text{DUT}(f) = \frac{\mathcal{L}_\text{DC}(f)}{V_\text{pp}^2\cdot 4\cdot \sin^2{(\pi f \tau )}}
\end{equation}

where $\mathcal{L}_\text{DUT}(f)$ is the phase noise PSD of the carrier signal, $\mathcal{L}_\text{DC}(f)$ is the measured, DC voltage noise PSD, $V_\text{pp}$ is the peak-to-peak voltage of the two 100 MHz tones heterodyned with one another, $f$ is the offset-frequency, and $\tau$ is the delay time of the optical DSH. 

With the DSH approach, the measured phase noise of the 111 GHz tone (shown in teal in Fig.~\ref{2}c.) reaches -148 dBc/Hz at 1 MHz offset frequency, about 20~dB lower than the measurement floor of the cross-correlation measurement at 1~MHz offset frequency. To reach the same improvement in the cross-correlation measurement we would have had to continuously average for $\approx$ 280 days. It should be noted that the measurement floor of the dual-tone DSH system was measured (plotted in grey in Fig.~\ref{3}c.) and is still the limiting factor in the millimeter-wave phase noise trace above 30 kHz, indicating a need for even lower noise measurement systems. This measurement floor comes from the uncorrelated noise along the electronic paths (i.e. noise from the photodetectors, amplifiers, and mixer).

\section*{Results}

\begin{figure}[]
  \includegraphics[width=\linewidth]{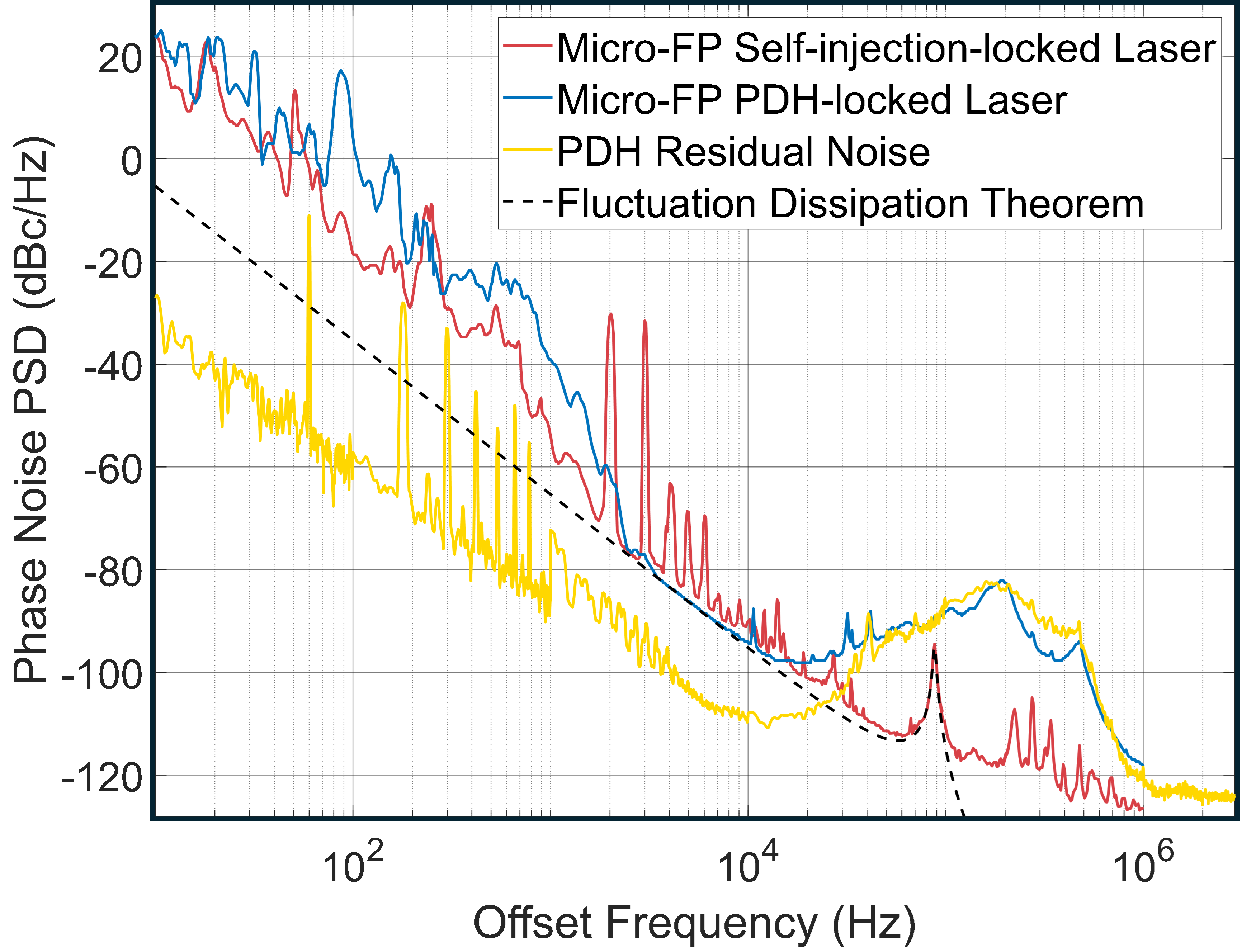}
\caption{Comparison phase noise plot of PDH-locking (blue) versus self-injection-locking (red) to the micro-FP. In yellow is the PDH-locking residual phase noise. In black dashed, the cavity limit predicted by fluctuation dissipation theorem. }
\label{3}
\end{figure}

As verified by the cross-correlation and DSH measurements (blue and teal curves of Fig. \ref{2}c), the millimeter-wave (111.45 GHz) resulting from the heterodyning of the dual-self-injection-locked lasers reaches phase noise of -110 dBc/Hz at 10 kHz offset and decreases as approximately $1/f^2$ to a value of -148 dBc/Hz at 1 MHz offset frequency. Comparing this to the out-of-loop phase noise of a single SIL laser (red, Fig.~\ref{2}c.), common mode rejection is responsible for at least 15 dB of correlated noise suppression in the heterodyned millimeter-wave signal across the entire spectrum. This common mode rejection is due to the fact that both lasers largely sample the same cavity noise \cite{Kessler2012}. Here the out-of-loop phase noise of the self-injection-locked laser was measured by a combination of direct heterodyne with a cavity-stabilized laser and single-tone DSH.

Figure \ref{3} illustrates the phase noise properties of a single laser relative to the cavity limit. The dashed black line of Fig. \ref{3} is the cavity limit defined by the fluctuation dissipation theorem \cite{Saulson1990}, which exhibits $1/f^3$ \cite{Numata2004} phase noise slope at offset frequencies less than 10 kHz, of the micro-FP cavity. In this same plot, we see that the phase noise of the SIL laser (red curve) approaches the cavity limit across a range of 1-100 kHz. However, it is notable that the phase noise of the mm-waves of Fig. \ref{2}c fall below this level, demonstrating that dual-self-injection-locking method enables millimeter-wave generation with phase noise beyond the cavity limit. Still, the correlated noise suppression is significantly less than what is predicted by theory ($\approx$ 64 dB, \cite{Liu2024APL}). Possible limitations could be the non-common fiber paths experienced by each laser in their independent self-injection-locking loops and the limited noise suppression of the free-running lasers with self-injection locking. One important note is that self-injection-locking does not provide an in-loop error signal to provide insight into the underlying locking limits whereas PDH-locking does. 

Previous results with a similar micro-FP that was free-space coupled predict a cavity thermal limit that is lower by about 15 dB \cite{Liu2024OPTICA}. 
This implies that the gluing of the two fiber collimators directly to the micro-FP cavity increased the mechanical damping, thereby increasing the cavity noise limit. This is accompanied by the appearance of a mechanical resonance near $\approx$ 88 kHz offset frequency which can be seen on the phase noise of the SIL laser (red curve of Fig. \ref{3}). Both the $1/f^3$ phase noise dependence and the mechanical resonance are well-described by a simple model of the thermal noise \cite{Saulson1990}. 

We also PDH-locked a narrow-linewidth fiber laser to the cavity to determine if the self-injection-locked laser was truly reaching the thermal noise limit of the cavity and self-injection loop. Here the phase noise of the PDH-locked laser was measured by direct heterodyne with a low-noise, cavity-stabilized laser. As shown by the blue curve of Fig. \ref{3}, in the 1-10 kHz range the noise of the PDH-locked laser overlaps both the SIL and cavity noise model. The residual laser noise, plotted in yellow, is well below the predicted cavity noise limit, suggesting the out-of-loop phase noise of the PDH-locked laser is not limited by locking electronics or the feedback loop. At frequencies below 1 kHz, the SIL and PDH-locked lasers diverge away from the $1/f^3$ slope characteristic of a cavity thermal limit. Likely, this excess noise is due to temperature, air pressure and vibrational fluctuations in the fiber and cavity. By comparing the red and blue curves of Fig. \ref{3}, we observe that at frequencies above 10 kHz, the SIL laser avoids the servo-bump drawback characteristic of PDH-lock-based systems, exhibiting as much as 40~dB lower phase noise than the PDH-locked laser. It should be noted that the diode lasers used here were not amenable to PDH-locking due to their larger linewidths. This demonstrates that passive self-injection locking enables the stabilization of noisier lasers than active PDH. 

\begin{figure}[]
  \includegraphics[width=\linewidth]{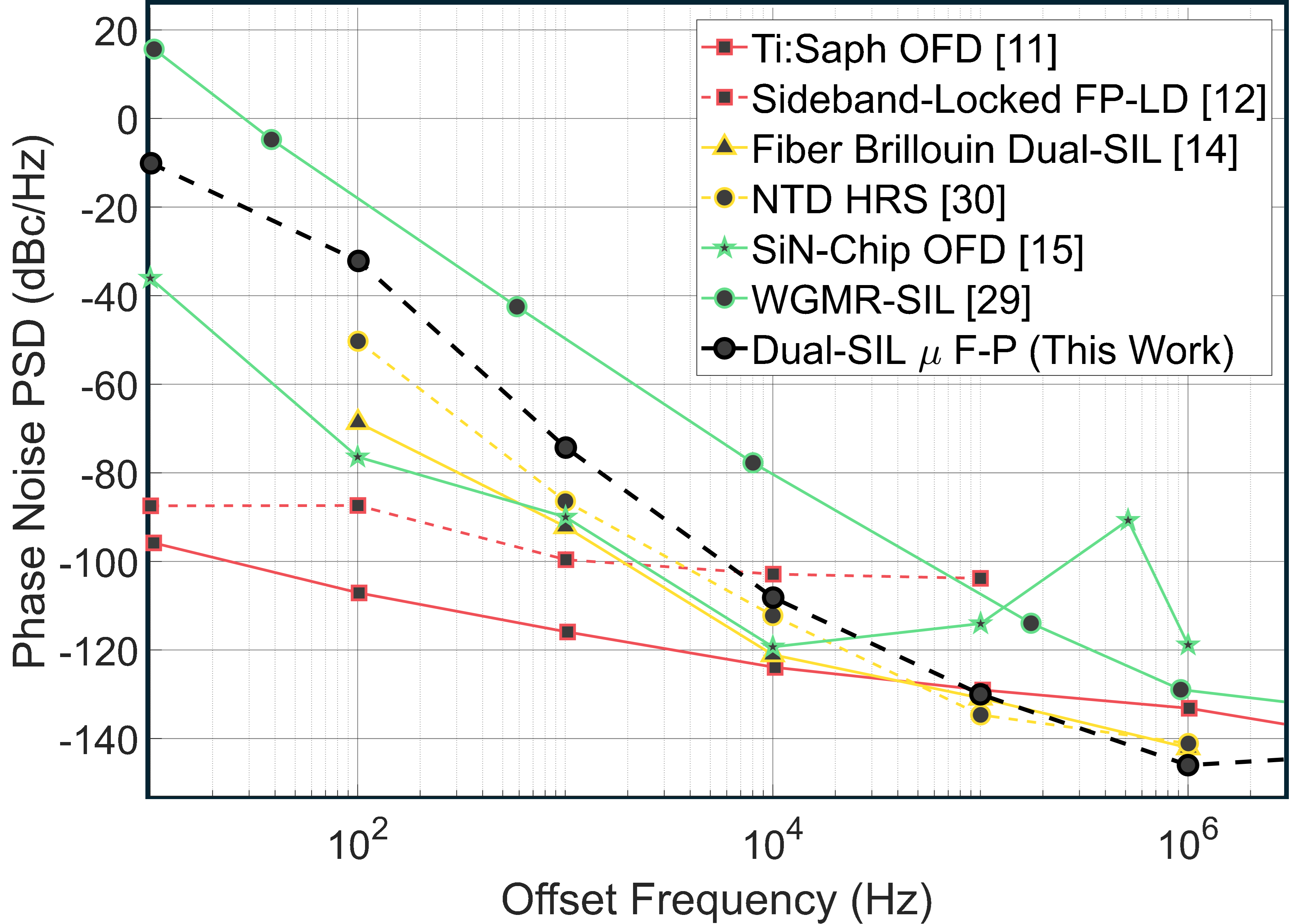}
\caption{Landscape of mm-wave generation techniques and phase noise, with lab scale architectures in red \cite{Fortier2016, Gan2015} (OFD~=~Optical frequency division, FP-LD = Fabry-Perot laser diode), small ($<$ 1 m$^3$) architectures in yellow \cite{Lia2023,Heffernan2024} (NTD HRS = Neutron-irradiated high-resistivity silicon), and integrable architectures in green \cite{Sun2024,Kittlaus2021} (WGMR~=~Whispering gallery mode resonator). }
\label{4}
\end{figure}

Lastly, we compare the results of our dual-SIL to micro-FP millimeter-wave generation to other state-of-the-art millimeter-wave generation techniques in Fig. \ref{4}. Here, it can be seen that integrable photonic architectures \cite{Sun2024,Kittlaus2021} for mm-wave generation are approaching bulk optic mm-wave phase noise performance at offset frequencies at and above 10 kHz. This remains an important offset frequency region of phase noise performance for radar and sensing applications. At higher offset frequencies, it is clear that this work, as well as work done with common-fiber spool cavities \cite{Heffernan2024} and photonic crystal oscillators \cite{Lia2023, Salek2023}, are performing as well as, if not better than bulk-optics \cite{Fortier2016, Gan2015}. At lower offset-frequencies, and hence longer time scales, more work must be done to stabilize these integrated systems to create long-millimeter-waves with long-term stability.

\section*{Conclusion and Outlook} 

In conclusion, our work presents a simple and compact photonic millimeter-wave source with phase noise comparable to state-of-the-art mm-wave sources, reaching -148~dBc/Hz at 1~MHz offset. Fabricating the micro-FP mirrors at wafer scale and using commercially available diode lasers provides a path to a unique manufacturable millimeter-wave source. On that point, an important viable use of these mm-wave sources is in mm-wave metrology. Having two of these dual-SIL synthesizers as references would allow one to cross-correlate faster at higher offset frequencies, by using direct mixing with the carrier signal. 

Furthermore, the lack of common-mode rejection in the dual-SIL system suggests an opportunity for still further phase noise gains. One immediate improvement is to remove as much excess fiber in the system as possible. We expect this to lead to a suppression in the out-of-loop SIL phase noise at offset frequencies less than 1 kHz, and consequently a reduction in the millimeter-wave phase noise. Other experiments involving dual-SIL with alternative cavities (i.e. microresonators, fiber cavities, whispering gallery mode resonators) as well as alternative coupling methods could also provide insight into the underlying mechanism of common-mode rejection in dual-SIL systems.

Lastly, coupling resonant light using glued fiber collimators is convenient, but it degrades the mechanical quality factor of the micro-FP cavity and increases its thermal noise. A future improvement to this system could involve hermetically-sealing the micro-FP with fiber collimators locked in place, close to the cavity mirrors. We expect such a configuration could provide and immediate improvement of the out-of-loop and millimeter-wave phase noise by 10-15 dB.

\bibliographystyle{ieeetr}
\bibliography{SILBeatMMBib}

\begin{thebibliography}{10}

\bibitem{Nagatsuma2016}
T.~Nagatsuma, S.~Hisatake, M.~Fujita, H.~H.~N. Pham, K.~Tsuruda, S.~Kuwano, and J.~Terada, ``Millimeter-{Wave} and {Terahertz}-{Wave} {Applications} {Enabled} by {Photonics},'' {\em IEEE Journal of Quantum Electronics}, vol.~52, pp.~1--12, Jan. 2016.

\bibitem{Lu2024}
H.-H. Lu, H.-M. Lin, C.-P. Wang, S.~T. Hayle, C.-Y. Li, X.-H. Huang, Y.-Y. Bai, K.~Okram, J.-M. Lu, Y.-C. Chung, and W.-W. Hsu, ``5 {G} new radio fiber-wireless converged systems by injection locking multi-optical carrier into directly-modulated lasers,'' {\em Communications Engineering}, vol.~3, p.~144, Oct. 2024.

\bibitem{Lu2023}
H.-H. Lu, C.-Y. Li, X.-H. Huang, C.-J. Lin, R.-D. Lin, Y.-S. Lin, Y.-S. Tang, and W.-C. Fan, ``A combined fibre/free-space-optical communication system for long-haul wireline/wireless transmission at millimetre-wave/sub-{THz} frequencies,'' {\em Communications Engineering}, vol.~2, p.~18, May 2023.

\bibitem{Dittmer2024}
J.~Dittmer, J.~Tebart, P.~Matalla, S.~Wagner, A.~Tessmann, A.~Bhutani, C.~Koos, A.~Stöhr, and S.~Randel, ``Comparison of electronic and optoelectronic signal generation for (sub-){THz} communications,'' {\em International Journal of Microwave and Wireless Technologies}, pp.~1--11, Nov. 2024.

\bibitem{Dong2022}
J.~Dong, Q.~Sun, Z.~Jiao, L.~Zhang, Z.~Yin, J.~Huang, J.~Yu, S.~Wang, S.~Li, X.~Zheng, and W.~Li, ``Photonics-enabled distributed {MIMO} radar for high-resolution {3D} imaging,'' {\em Photonics Research}, vol.~10, p.~1679, July 2022.

\bibitem{Yi2023}
L.~Yi, Y.~Li, and T.~Nagatsuma, ``Photonic {Radar} for {3D} {Imaging}: {From} {Millimeter} to {Terahertz} {Waves},'' {\em IEEE Journal of Selected Topics in Quantum Electronics}, vol.~29, pp.~1--14, Sept. 2023.

\bibitem{Bai2023}
W.~Bai, P.~Li, X.~Zou, N.~Zhong, W.~Pan, L.~Yan, and B.~Luo, ``Photonic super-resolution millimeter-wave joint radar-communication system using self-coherent detection,'' {\em Optics Letters}, vol.~48, p.~608, Feb. 2023.

\bibitem{Zhao2024}
Y.~Zhao, A.~A. Abbas, M.~Sakaki, G.~Bramlage, G.~Delaittre, N.~Benson, T.~Kaiser, and J.~C. Balzer, ``3d printed sub-terahertz photonic crystal for wireless passive biosensing,'' {\em Communications Engineering}, vol.~3, p.~69, May 2024.

\bibitem{Chen2024}
Y.~Chen, J.~Yuan, and J.~Tang, ``A high precision vital signs detection method based on millimeter wave radar,'' {\em Scientific Reports}, vol.~14, p.~25535, Oct. 2024.

\bibitem{Zhang2023}
J.~Zhang, R.~Xi, Y.~He, Y.~Sun, X.~Guo, W.~Wang, X.~Na, Y.~Liu, Z.~Shi, and T.~Gu, ``A {Survey} of {mmWave}-{Based} {Human} {Sensing}: {Technology}, {Platforms} and {Applications},'' {\em IEEE Communications Surveys \& Tutorials}, vol.~25, no.~4, pp.~2052--2087, 2023.

\bibitem{Fortier2016}
T.~M. Fortier, A.~Rolland, F.~Quinlan, F.~N. Baynes, A.~J. Metcalf, A.~Hati, A.~D. Ludlow, N.~Hinkley, M.~Shimizu, T.~Ishibashi, J.~C. Campbell, and S.~A. Diddams, ``Optically referenced broadband electronic synthesizer with 15 digits of resolution,'' {\em Laser \& Photonics Reviews}, vol.~10, pp.~780--790, 9 2016.

\bibitem{Gan2015}
L.~Gan, J.~Liu, F.~Li, and P.~K.~A. Wai, ``An {Optical} {Millimeter}-{Wave} {Generator} {Using} {Optical} {Higher} {Order} {Sideband} {Injection} {Locking} in a {Fabry}–{Pérot} {Laser} {Diode},'' {\em Journal of Lightwave Technology}, vol.~33, pp.~4985--4996, Dec. 2015.

\bibitem{Testumoto2021}
T.~Tetsumoto, T.~Nagatsuma, M.~E. Fermann, G.~Navickaite, M.~Geiselmann, and A.~Rolland, ``Optically referenced 300 {GHz} millimetre-wave oscillator,'' {\em Nature Photonics}, vol.~15, pp.~516--522, July 2021.

\bibitem{Heffernan2024}
B.~M. Heffernan, J.~Greenberg, T.~Hori, T.~Tanigawa, and A.~Rolland, ``Brillouin laser-driven terahertz oscillator up to 3 {THz} with femtosecond-level timing jitter,'' {\em Nature Photonics}, vol.~18, pp.~1263--1268, Dec. 2024.

\bibitem{Sun2024}
S.~Sun, B.~Wang, K.~Liu, M.~W. Harrington, F.~Tabatabaei, R.~Liu, J.~Wang, S.~Hanifi, J.~S. Morgan, M.~Jahanbozorgi, Z.~Yang, S.~M. Bowers, P.~A. Morton, K.~D. Nelson, A.~Beling, D.~J. Blumenthal, and X.~Yi, ``Integrated optical frequency division for microwave and mmwave generation,'' {\em Nature}, vol.~627, pp.~540--545, 3 2024.

\bibitem{Groman2024}
W.~Groman, I.~Kudelin, A.~Lind, D.~Lee, T.~Nakamura, Y.~Liu, M.~L. Kelleher, C.~A. McLemore, J.~Guo, L.~Wu, W.~Jin, K.~J. Vahala, J.~E. Bowers, F.~Quinlan, and S.~A. Diddams, ``Photonic millimeter-wave generation beyond the cavity thermal limit,'' {\em Optica}, vol.~11, p.~1583, Nov. 2024.

\bibitem{Hadley1986}
G.~Hadley, ``Injection locking of diode lasers,'' {\em IEEE Journal of Quantum Electronics}, vol.~22, no.~3, pp.~419--426, 1986.

\bibitem{Jin2022}
N.~Jin, C.~A. McLemore, D.~Mason, J.~P. Hendrie, Y.~Luo, M.~L. Kelleher, P.~Kharel, F.~Quinlan, S.~A. Diddams, and P.~T. Rakich, ``Micro-fabricated mirrors with finesse exceeding one million,'' {\em Optica}, vol.~9, p.~965, 9 2022.

\bibitem{Liu2024OPTICA}
Y.~Liu, N.~Jin, D.~Lee, C.~McLemore, T.~Nakamura, M.~Kelleher, H.~Cheng, S.~Schima, N.~Hoghooghi, S.~Diddams, P.~Rakich, and F.~Quinlan, ``Ultrastable vacuum-gap {Fabry}–{Perot} cavities operated in air,'' {\em Optica}, vol.~11, p.~1205, Sept. 2024.

\bibitem{Li2021}
B.~Li, W.~Jin, L.~Wu, L.~Chang, H.~Wang, B.~Shen, Z.~Yuan, A.~Feshali, M.~Paniccia, K.~J. Vahala, and J.~E. Bowers, ``Reaching fiber-laser coherence in integrated photonics,'' {\em Optics Letters}, vol.~46, p.~5201, 10 2021.

\bibitem{LiuBlumenthal2022}
K.~Liu, N.~Chauhan, J.~Wang, A.~Isichenko, G.~M. Brodnik, P.~A. Morton, R.~O. Behunin, S.~B. Papp, and D.~J. Blumenthal, ``36 {Hz} integral linewidth laser based on a photonic integrated 4.0 m coil resonator,'' {\em Optica}, vol.~9, p.~770, July 2022.

\bibitem{ChengYale2024}
H.~Cheng, C.~Xiang, N.~Jin, I.~Kudelin, J.~Guo, M.~Heyrich, Y.~Liu, J.~Peters, Q.-X. Ji, Y.~Zhou, K.~J. Vahala, F.~Quinlan, S.~A. Diddams, J.~E. Bowers, and P.~T. Rakich, ``Harnessing micro-{Fabry}-{Perot} reference cavities in photonic integrated circuits,'' Oct. 2024.
\newblock arXiv:2410.01095 [physics].

\bibitem{Rubiola2010}
E.~Rubiola and F.~Vernotte, ``The cross-spectrum experimental method,'' Mar. 2010.
\newblock arXiv:1003.0113 [physics].

\bibitem{Kwon2017}
D.~Kwon, C.-G. Jeon, J.~Shin, M.-S. Heo, S.~E. Park, Y.~Song, and J.~Kim, ``Reference-free, high-resolution measurement method of timing jitter spectra of optical frequency combs,'' {\em Scientific Reports}, vol.~7, p.~40917, Jan. 2017.

\bibitem{Kessler2012}
T.~Kessler, T.~Legero, and U.~Sterr, ``Thermal noise in optical cavities revisited,'' {\em Journal of the Optical Society of America B}, vol.~29, p.~178, Jan. 2012.

\bibitem{Saulson1990}
P.~R. Saulson, ``Thermal noise in mechanical experiments,'' {\em Phys. Rev. D}, vol.~42, pp.~2437--2445, Oct 1990.

\bibitem{Numata2004}
K.~Numata, A.~Kemery, and J.~Camp, ``Thermal-noise limit in the frequency stabilization of lasers with rigid cavities,'' {\em Phys. Rev. Lett.}, vol.~93, p.~250602, Dec 2004.

\bibitem{Liu2024APL}
Y.~Liu, D.~Lee, T.~Nakamura, N.~Jin, H.~Cheng, M.~L. Kelleher, C.~A. McLemore, I.~Kudelin, W.~Groman, S.~A. Diddams, P.~T. Rakich, and F.~Quinlan, ``Low-noise microwave generation with an air-gap optical reference cavity,'' {\em APL Photonics}, vol.~9, 1 2024.

\bibitem{Kittlaus2021}
E.~A. Kittlaus, D.~Eliyahu, S.~Ganji, S.~Williams, A.~B. Matsko, K.~B. Cooper, and S.~Forouhar, ``A low-noise photonic heterodyne synthesizer and its application to millimeter-wave radar,'' {\em Nature Communications}, vol.~12, p.~4397, 7 2021.

\bibitem{Lia2023}
E.~Lia, I.~Ghosh, S.~M. Hanham, B.~Walter, F.~Bavedila, M.~Faucher, A.~P. Gregory, L.~Jensen, J.~Buchholz, H.~Fischer, U.~Altmann, and R.~Follmann, ``Novel mm-{Wave} {Oscillator} {Based} on an {Electromagnetic} {Bandgap} {Resonator},'' {\em IEEE Microwave and Wireless Technology Letters}, vol.~33, pp.~863--866, June 2023.

\bibitem{Salek2023}
M.~Salek, S.~M. Hanham, and A.~P. Gregory, ``High- \textit{{Q}} 100 {GHz} {Photonic} {Crystal} {Resonator} {Fabricated} {From} a {Cyclic} {Olefin} {Copolymer},'' {\em IEEE Microwave and Wireless Technology Letters}, vol.~33, pp.~279--282, Mar. 2023.

\end{thebibliography}

\medskip
\begin{footnotesize}

\noindent\textbf{Corresponding authors}: \href{mailto:william.groman@colorado.edu}{william.groman@colorado.edu} and \href{mailto:scott.diddams@colorado.edu}{scott.diddams@colorado.edu}

\noindent \textbf{Funding}: 
This research was supported by DARPA GRYPHON program (HR0011-22-2-0009) and NIST % National Aeronautics and Space Administration
(80NM0018D0004).

\noindent \textbf{Acknowledgments}: 
Commercial equipment and trade names are identified for scientific clarity only and does not represent an endorsement by NIST.

\noindent \textbf{Author Contributions}:

W.G., N.J., H.C., F.Q., P.R., and S.A.D. conceived the experiment and supervised the project.
W.G., and S.A.D. wrote the paper with input from all authors.
W.G., N.J., and H.C.,  built the locking/tuning mechanisms and performed the mm-wave experiment.
M.H. provided the stable cavity-locked optical reference for measuremnts.
D.M., M.H., Y.L., C.A.M., and A.L. provided input regarding the cavity and millimeter-wave measurements. 
N.J. and H.C. built the micro-FP cavity and provided information regarding the cavity. 
%L.W., prepared the DFB laser butterfly packages for the experiment.
%Q.-X.J., J.G., W.J., L.W., and C.X. prepared the microcomb and spiral resonators for the experiment.
%M.L.K., and F.Q. built the F-P cavity
%D.L., T.N., C.A.M., Y.L., and F.Q. provided optically derived microwave reference and aided in the microwave phase noise measurement system.
%J.B. and J.C.C. provided MUTC detectors.
All authors contributed to discussion of the results.

\noindent \textbf{Disclosures}: 

The authors declare no conflicts of interest.

%\noindent \textbf{Data and materials availability}:  

%All data for the figures in this manuscript is available at \url{}.

\end{footnotesize}

\end{document}